\documentstyle[epsf]{mn}
\def \mpc       {{\rm\ Mpc}}

\def \etal      {\hbox{et al.}}
\def \dln       {\hbox{\rm d$\,$ln }}

\def \Ho        {{\rm\ H_{o}}}
\def \kmsmpc    {{\rm\ km\ s^{-1}\ Mpc^{-1}}}
\def \kev       {{\rm\ keV}}
\def \msol      {{\rm M}_\odot}

\def \hinv      {\hbox{$\, h^{-1}$} }

\def \hfiftyinv {\hbox{$\, h_{50}^{-1}$} }
\def \ergs      { \hbox{$\,$ erg s$^{-1}$} }

\def \se        {\!=\!}

\def \ssimeq    {\! \simeq \!}
\def \sequiv    {\! \equiv \!}
\def \spropto   {\! \propto \!}
\def\\{\hfil\break}
\def\spose#1{\hbox to 0pt{#1\hss}}
\def\lta{\mathrel{\spose{\lower 3pt\hbox{$\mathchar"218$}}
     \raise 2.0pt\hbox{$\mathchar"13C$}}}
\def\gta{\mathrel{\spose{\lower 3pt\hbox{$\mathchar"218$}}
     \raise 2.0pt\hbox{$\mathchar"13E$}}}

\def\lesssim{\mathrel{\hbox{\rlap{\hbox{\lower4pt\hbox{$\sim$}}}\hbox{$<$}}}}
\def\gtrsim{\mathrel{\hbox{\rlap{\hbox{\lower4pt\hbox{$\sim$}}}\hbox{$>$}}}}
\newcount\itemno
\def \ino         { \the\itemno\global\advance\itemno by 1 }
\def \xray {\hbox{X--ray} }

\def \rhocrit {\hbox{$\rho_c$}}

\def \lognlogs {\hbox{logN--logS} }
\def \Mstar {\hbox{$M_{\star}$}}
\def \LxT {\hbox{$L_X$--$T$} }

\title{\bf Constraints on $\Omega_0$ and Cluster Evolution using the 
ROSAT LogN--LogS }
\author[B. Mathiesen and A. Evrard]
    {B. Mathiesen$^1$ and A. E. Evrard$^{1,2,3}$ \\
    $^1$ Physics Dep't, 2071 Randall Lab, University of Michigan, Ann Arbor,
MI 48109--1120 USA \\
   $^2$ Institut d'Astrophysique, 98bis Blvd Arago, 75014 Paris, France \\
   $^3$ Max--Planck--Institut f\"ur Astrophysik, Karl--Schwarz.--Str. 1,
Garching bei M\"unchen, Germany  \\
    bfm@umich.edu \\ evrard@umich.edu}

\date{\today}

\pubyear{1997}

\begin{document}

\maketitle

%change title for purpose of shorter page heading

\begin{abstract}
We examine the likelihoods of different cosmological models and 
cluster evolutionary histories by comparing semi-analytical 
predictions of X-ray cluster number counts to
observational data from the ROSAT satellite.  
We model cluster abundance as a function of mass and
redshift using a Press-Schechter distribution, and 
assume the temperature $T(M,z)$ and bolometric luminosity $L_X(M,z)$ 
scale as power laws in mass and epoch, in order to construct 
expected counts as a function of X--ray flux.  The $L_x-M$ scaling is 
fixed using the local luminosity function while the degree 
of evolution in the X-ray luminosity with redshift 
$L_X \spropto (1+z)^s$ is left open, with 
$s$ an interesting free parameter which we investigate.  
We examine open and flat cosmologies with initial, 
scale--free fluctuation spectra having indices 
$n = 0$, $-1$ and $-2$.  An independent constraint arising from
the slope of the luminosity--temperature relation strongly favors 
the $n \se -2$ spectrum. 

The expected counts demonstrate a strong dependence on $\Omega_0$ and 
$s$, with lesser dependence on $\lambda_0$ and $n$.  Comparison with
the observed counts reveals a ``ridge'' of acceptable models in 
$\Omega_0 - s$ plane, roughly following the relation $s \sim 6 \Omega_0$ 
and spanning low-density models with a small degree of evolution
to $\Omega = 1$ models with strong evolution.  Models with moderate
evolution are revealed to have a strong lower limit of $\Omega_0 \gtrsim 0.3$,
and low-evolution models imply that $\Omega_0 < 1$ 
at a very high confidence level.  We suggest observational tests for 
breaking the degeneracy along this ridge, and discuss implications 
for evolutionary histories of the intracluster medium.  

\end{abstract}

%taken from most recent MNRAS headings index
\begin{keywords}
cosmology: theory --
cosmology: observation --
clusters: galaxies: general --
dark matter
\end{keywords} 

\section{Introduction}

Rich galaxy clusters are the youngest virialized objects extant, and 
as such, they provide a unique source of information about our
universe.  Observations of clusters in the X-ray band provide 
useful information about large-scale structure and galaxy formation.  
Detailed X-ray images provide information
about the structure of individual clusters, and surveys 
such as the ROSAT all-sky survey (RASS) provide data on the 
general cluster population.  In particular, counts of clusters
as a function of their \xray flux --- the logN--logS relation --- 
provide an avenue for exploring cosmic evolution of clusters.  
Recently, several groups have pushed the \lognlogs relation to fluxes 
nearly an order of magnitude fainter than previous determinations 
(Rosati \etal 1995; Rosati \& Della Ceca 1997; Jones \etal 1997).  In this 
paper, we compare these data to predictions 
of viable cosmological and evolutionary models for clusters, with the
goal of defining the range of combined cosmology and \xray evolution 
consistent with the current data.  Because the data 
underconstrain the theory, a range of degenerate models emerge, but 
these can be distinguished by upcoming observational 
tests of the high redshift cluster population.

In order to proceed, we assume the distribution of clusters as a
function of mass and redshift is accurately described by a 
Press-Schecter (1974) abundance function.  To investigate the behavior
of a wide class of structure formation models, we 
employ scale--free initial Gaussian perturbations 
with spectral indices $n$ equal to 0, $-1$, and $-2$ in the cluster 
mass regime.  (The effect of curvature in the spectrum is discussed
below.) We allow $\Omega_0$ to vary and investigate both 
open models without a cosmological constant 
($\lambda_0 \equiv \Lambda/3\Ho^2 = 0$) and flat models 
where $\Omega_0 + \lambda_0 = 1$.  We assume a value of 
$\Ho \se 50 \kmsmpc$ throughout this work, in accordance 
with the traditional treatment of \xray data, but most of our analysis
is independent of this choice.  

The Press-Schecter formalism predicts the number density of collapsed dark
matter halos as a function of their mass and redshift. Since what we observe
is the band-limited X-ray luminosity ($L_X(E)$), we need to find a way to
relate these quantities to one another. Rather than adopt a specific
``microphysical'' model to describe the relationship between mass and
bolometric luminosity, we assume that it can be adequately described by
a power law in the mass regime of interest, and then fit the free parameters
of the power law using the local X-ray luminosity function (XLF). The
emissivity of the gas is modeled as a thermal Bremsstrahlung
spectrum, which is integrated to determine the fraction of
the bolometric luminosity falling in the appropriate energy band.
We use the  ROSAT Brightest Cluster Sample (BCS) compiled by Ebeling \etal
(1997) to constrain the parameters of the fit.

Since X-ray luminosity is proportional to the squared density of 
ions in the intracluster medium (ICM), and since the early universe
was denser than it is today, it is reasonable to expect that
distant clusters may be stronger X-ray emitters despite
having lower overall masses.  Arguments based on self--similarity 
predict evolution in the bolometric X-ray luminosity of the form 
$L_X \propto M^{4/3} (1+z)^{7/2}$ (Kaiser 1986), implying strong 
positive evolution in the luminosity of objects at a fixed mass.  
However, self--similarity is not particularly well justified either
obervationally or theoretically; models which invoke 
a minimum central entropy in the cluster gas fare better in many
respects (Evrard \& Henry 1991; Kaiser 1991).  

The issue of evolution in the XLF has been 
hotly pursued among observers; whether or not any evolution exists,
however, still seems an open question.  Early work on the Extended 
Medium Sensitivity Survey (EMSS) cluster sample by Henry et al. (1992) 
and Gioia et al. (1990a) found intriguing evidence for evolution in the
cluster population.  These results sparked interest but were
preliminary analyses that suffered from occasional misclassifications
and the vagaries of small-number statistics. These same clusters were
recently reanalyzed with new X-ray observations by Nichol et al. (1997),
who found that the data were consistent 
with no evolution in the cluster population out to redshifts of about 0.3.
The BCS set of 199 clusters mentioned above also shows no
evidence of evolution in the XLF at low redshifts.
More distant observations include those of Castander et al. (1995), Luppino
\& Gioia (1995), and Collins \etal (1997).
Luppino and Gioia have been collecting high-redshift
clusters from the EMSS sample and finding number densities higher than the
standard $\Omega \se  1$ CDM model predicts, which could indicate either a
low-density  universe or the existence of a strong degree of evolution.
Castander \etal have found a dearth of high-redshift clusters relative to
a simple no-evolution prediction based on integration of the local XLF.
Collins et al., however, compiled a sample of high-reshift clusters
from the SHARC (Serindipitous High-redshift Archival ROSAT Cluster) survey,
and found significantly higher numbers at $z > 0.3$.

Given the lack of consensus in this debate, 
we allow for freedom in how luminosity scales with redshift at a fixed
mass by writing $L_X \propto (1+z)^s$.  The cosmological models described
above are then examined under varying degrees of
evolution in the X-ray luminosity, yielding four interesting free
parameters for the models considered in this paper: $\Omega_0$, $\lambda_0$,
$n$, and $s$.  
Armed with the above assumptions, we extrapolate the local cluster
abundances to high redshifts and predict the surface densities of clusters
at low fluxes.  Comparing predictions to recent observational determinations
of the cluster \lognlogs (Rosati \& Della Ceca 1997; Jones \etal 1997) allows 
us to make statements about the relative probablilities of the models.  

In \S 2, we describe the mathematical model used and the
process of fixing a subset of parameters with local observations.
In \S 3, we describe our method of predicting the \lognlogs 
and the status of current measurements of this quantity.  We also 
discuss the likelihoods of individual models and the cosmological constraints
which can be obtained from analysis of the \lognlogs
alone.  In \S 4, we go on to explore further ways of discriminating
among cosmologies, by making use of the luminosity--temperature 
relation and redshift distributions of flux limited samples.  
Finally, in section 5, we sum up our results and
suggest future directions for theory and observation in this rich field.  

\section{Fixing the Model}

\subsection{Theoretical Framework}

The method described here is similar to that put forth in Evrard \& Henry 
(1991, hereafter referred to as EH91).
The first step in predicting the number of observable X-ray clusters at a
given flux limit is to model the distribution of these objects.  Given that
X-ray clusters correspond to virialized dark matter (DM) halos, we 
assume that the population of these objects is well-described by a 
Press-Schechter distribution (e.g., Lacey and Cole 1993)
\begin{equation}
\frac{dn(M,z)}{dM} = -\sqrt{\frac{2}{\pi}}\frac{\bar{\rho}(z)}{M^2} 
\frac{d\ln\sigma} {d\ln M}\nu(M,z)\exp\left[\frac{-\nu^2(M,z)}{2}\right]
\end{equation}
where $dn(M,z)$ is the number density of collapsed halos in the mass range
$[M,M+dM]$ and $\bar{\rho}(z)$ is the mean background density at
redshift $z$.  The normalized fluctuation amplitude $\nu(M,z)$ 
is defined as $\delta_{c0}(z)/\sigma(M)$, where
$\sigma(M)$ is the variance of the fluctuation spectrum filtered on 
mass scale $M$ and $\delta_{c0}(z)$ is
the linearly evolved overdensity of a perturbation that has collapsed and
virialized at a redshift $z$ (see Appendix A).  
We assume a scale--free power spectrum 
$P(k)\propto k^n$, so the variance can be written 
\begin{equation}
\sigma(M) = \sigma_{15} M^{-\alpha}
\end{equation}
where the subscript indicates that the normalization is to a mass of 
$10^{15} \hfiftyinv M_\odot$ (with $h_{50} \se \Ho/50 \kmsmpc$) 
and $\alpha \se (n+3)/6$.  
Connection to the conventional normalization $\sigma_8$ within
$8 \hinv \mpc$ spheres is straightforward
\begin{equation}
\sigma_{15} = [\frac{4\pi}{3}(16\hfiftyinv)^3\rhocrit\Omega_0/10^{15}
\hfiftyinv]^\alpha \sigma_8 = [1.19 \Omega_0]^\alpha \sigma_8
\end{equation}
where $\rhocrit \se 3\Ho^2/8\pi G$ is the critical density.  
The power spectrum normalization deduced from cluster abundances 
follows the empirical fitting function
\begin{equation}
\sigma_8 = c_1 \Omega_0^{c_2}
\end{equation}
where there is good, but not exact, agreement in the literature on the
values of $c_1$ and $c_2$ (White, Efstathiou \& Frenk 1993; Viana
\& Liddle 1996; Eke, Cole \& Frenk 1996).  For our calculations,
we use $c_1 = 0.60$, 
\begin{equation}
c_2(\Omega_0) = 0.36 + 0.31\Omega_0 - 0.28\Omega_{0}^{2}
\end{equation}
for open models with $\lambda_0 = 0$, and
\begin{equation}
c_2(\Omega_0) = 0.59 - 0.16\Omega_0 + 0.06\Omega_{0}^{2}
\end{equation}
for models with $\Omega_0 + \lambda_0 = 1$.  These fitting functions were
taken from Viana \& Liddle (1996).  There is a random uncertainy on $\sigma_8$
of approximately $\pm^{37\%}_{27\%}$, the main component of which results from
the cosmic variance in the local cluster population. Since this is taken into
account in a different manner later in the paper (section 3.3), we treat
the above value as exact. 

We assume that the bolometric X-ray luminosity of clusters follows a power 
law in mass and redshift 
\begin{equation}
L_X = L_{15} \ M^p \ (1+z)^s 
\label{LMz_relation}
\end{equation}
over a range of $10^{13}$ to $10^{16} M_\odot$ in mass 
and $z \lta 2$ in redshift.  Here, and throughout the paper unless 
specified otherwise, the mass $M$ is in units of $10^{15} \hfiftyinv \msol$.  
Although assuming that this simple mass dependence holds over three orders of
magnitude may seem unreasonable, the abundance of objects drops quite
sharply outside a central range of about 1.5 decades for the flux limits that
we are considering.  High-mass objects are rare at any epoch, and low-mass
objects quickly become invisible at larger redshifts; thus, any
deviations from a power law outside this range will have little effect 
on our predictions.  The intrinsic luminosity at fixed mass increases 
with redshift for $s > 0$.  For reasons which will be made apparent, we focus 
attention on such ``positive luminosity evolution'' models, though it is 
straightforward to extrapolate our results to models with negative
luminosity evolution.  

What values of the parameters $p$ and $s$ are expected?  
On dimensional grounds, we can write a scaling relation 
for the bolometric luminosity as 
\begin{equation}
L_X \propto \int dV \rho^2 T^{1/2} \propto M^{4/3} (1+z)^{7/2} I(M,z) 
\end{equation}
where the virial theorem $T \propto M^{2/3} (1+z)$ and the assumption
of clusters as regions of fixed overdensity $\int dV \rho^2 \sim M
\bar{\rho}(z) \sim M (1+z)^3$ with constant gas fraction 
is used to produce the scalings on the
right hand side.  Here $I(M,z)$ is a form factor which retains the 
information on the mean internal density and temperature profiles of
the clusters.  Kaiser (1986) derived the above scaling under the
assumption of self-similarity of the cluster population across both
mass and epoch, so that $I(M,z) = const$.  Although self--similarity 
may apply to the cluster popultion, it requires rather restrictive 
conditions --- gravitational shock heating should be the 
dominant heating mechanism, cooling unimportant, variations in gas 
fraction must be small, clusters must have similar internal structure, and so
on. EH91 presented empirical evidence against self-similar scaling in 
mass for $\Omega \se 1$ models from the shape of the luminosity function.  
(We return to this point below.)   Kaiser (1991) and EH91 presented 
alternative models which invoked constant entropy either throughout the 
cluster gas or in the central core, respectively.  The constant core 
entropy models of EH91 yield $p \se 11/6$ and $s \se 11/4$.  

The approach taken here is to let the observations dictate appropriate 
values of $p$ and $s$.  The value of $p$ then reflects a 
composite mean description summarizing the mass sensitivity of 
cluster internal structure, gas fraction, cooling flows, 
efficiency of star formation and other microphysics.  Variance in this
relation is discussed below.  The
evolution parameter $s$ encompasses time-dependent phenomena 
such as the changing overall density of the universe, the
efficiency of radiative cooling, and
heating of the ICM via gravitational collapse or supernova injection.
Bower (1997) discusses the connection between
cluster entropy and the evolution parameter.
The value $s \se 11/4$ indicates that these processes are balanced; higher 
values indicate that heating mechanisms are dominant, while lower values
indicate that cooling mechanisms are dominant.  

The above discussion pertains to the bolometric cluster luminosity. 
In practice, the \xray luminosity within some range of photon
energies, denoted by energy band $E$, 
is required to connect with the observational data.  The observed 
luminosity $L_X(E)$ is a fraction of the bolometric 
\begin{equation}
L_X(E) = L_X \  f_E[T(M,z),z]
\end{equation}
where the factor $f_E[T(M,z),z]$ is found by numerically integrating the
Bremsstrahlung emissivity over the proper energy range.  As in EH91, 
we use an approximation to the Gaunt factor of $g(E,kT) = 0.9(E/kT)^{-0.3}$
in this calculation.
The temperature of the cluster is related to the mass according to the
equation $kT$(keV) = $3.96 M^{2/3}(1+z)$, a scaling law that is well
supported by three-dimensional hydrodynamic simulations (Evrard
1990a,b; Evrard, Metzler \& Navarro 1996).\footnote{Although using this
equation assumes that the cluster gas is fully virialized, the value of
the band fraction is not strongly dependent on temperature and minor deviations
from equilibrium will not significantly affect the results of this
paper.}
Since the cluster temperature and emitted photon energy both scale as 
$(1+z)$ the redshift dependence for a fixed received energy band 
drops out and the X-ray luminosity fraction
can be easily tabulated as a function of $M$ or $T$ alone.    
We use an energy range of $0.1-2.4 \kev$ when comparing our model to the local 
luminosity function, and $0.5-2.0 \kev$ when predicting cluster number counts. 

\subsection{Constraints from the local luminosity} 

\begin{figure}
\epsfxsize 8.5cm
\epsfysize 8.5cm
\epsfbox{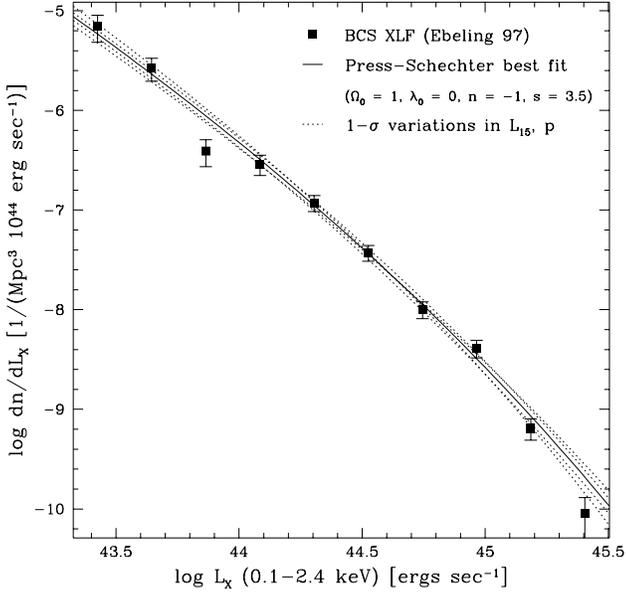}
\caption{The BCS XLF of Ebeling \etal (1997).  A change in the slope of the
abundance function is evident around an X--ray luminosity of $3 \times 10^{44}
\ergs$.  The solid line is the least-squares fit to a Press-Schechter
form as described in the text.  The dotted lines are the curves that correspond
to $L_{15}$ and $p$ values at the vertices of the one-sigma error ellipse
derived from the covariance matrix of the fit.  The BCS XLF extends over a
larger luminosity range than is shown here, but the additional data points
do not help to constrain the parameters further.}
\label{BCSdat}
\end{figure}

The free parameters of our mass-luminosity model are $L_{15}$, $p$ and $s$.
Since the extent of redshift evolution is still in question,
we allow $s$ to vary in the range $0.0 \leq s \leq 6.0$.
$L_{15}$ and $p$ can then be determined by inverting 
equation~(\ref{LMz_relation}) and expressing the Press-Schechter
abundance as $n(L_X,z)dL_X$. The resulting function is then fit to the BCS
XLF of Ebeling \etal (1997), using an average redshift of 0.1 for the entire
sample.  Figure~\ref{BCSdat} shows the results of fitting a standard
CDM--like model ($\Omega \se 1$, $n \se -1$, $\lambda_0 \se 0$, and 
$s \se 3.5$) to these data, with resulting least-squares fit values of
$p = 3.38 \pm 0.17$ and 
$L_{15} = 2.50 \pm 0.16 \times 10^{44} \, h_{50}^{-2} \ergs$. 

\begin{figure}
\epsfxsize 8.5cm
\epsfysize 8.5cm
\epsfbox{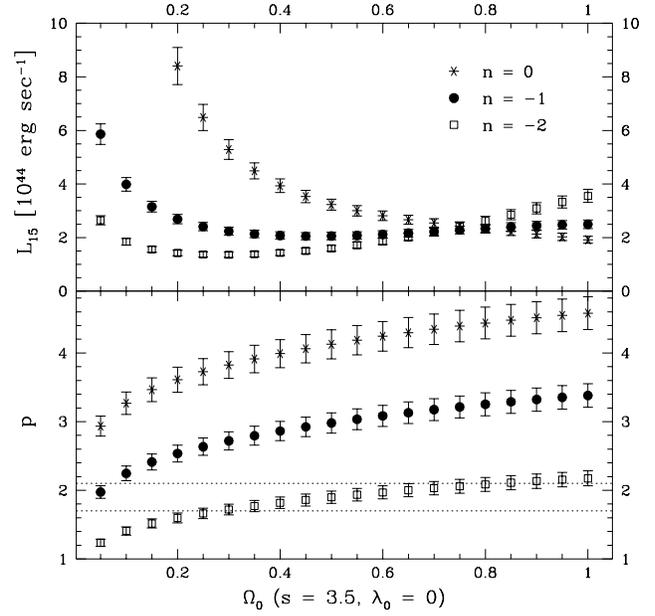}
\caption{The best-fit values of $L_{15}$ and $p$ for open cosmological models,
assuming an evolution parameter $s \se 3.5$.  The error bars correspond to
the formal 1-$\sigma$ uncertainty given by the diagonal elements of the
covariance matrix.  The overall shape of the curves and the
level of uncertainty are similar in models which have different $s$ or
include a cosmological constant.  The dotted lines are the limits on $p$
that result if the observed luminosity-temperature relation is adopted as
an additional constraint; this topic will be discussed in detail in \S
4.1.  The numerical value of 
$L_{15}$, defined as the luminosity at mass $10^{15} \hfiftyinv
\msol$, scales as $h_{50}^{-2}$.
}  
\label{l15pdat}
\end{figure}

Values of $L_{15}$ and $p$ for any given cosmological model can be similarly 
determined to roughly 10\% and 5\% accuracy, respectively.
Figure~\ref{l15pdat} displays the variation in $L_{15}$ and $p$ with
$\Omega_0$ and $n$ at a constant value of $s \se 3.5$.
The level of accuracy displayed is representative of all models.  
We analyze each model individually and employ its best-fit values for these
parameters in subsequent calculations.  All models provide good fits to the
BCS XLF, with reduced chi-squared parameters of
$\langle\chi^2\rangle \approx 1.2$ over seven degrees of freedom.

The behavior of $p$ in Figure 2 can be understood
by examining the Press-Schechter abundance function,  
equation~(1), and the variation of $\sigma_8$ with $\Omega_0$ in open
and flat models, equations~(5) and (6).  For a given set of cosmological 
parameters the form of the underlying mass function is completely fixed.
For masses smaller than $\Mstar \!\equiv\! 
(\sqrt{2}\sigma_{15}/\delta_{c0})^{1/\alpha}$,
equation~(1) behaves like a power law: $n(M) \sim M^{-2+\alpha}$.  If 
$M > \Mstar$, however, we have $n(M) \sim \exp[-(M/\Mstar)^{2\alpha}]$.
As $n$ increases, the exponential decay gets stronger and the power law gets
more shallow, creating a stronger bend in the mass function at the transition
region.  In making the transformation from $n(M,z)$ to $n(L_X,z)$, we find
that $M \sim L_X^{1/p}$, so a larger $p$ will stretch out the luminosity
function more and produce a shallower bend at $L_X(\Mstar)$.  Indeed, this is
what we see; larger values of $n$ systematically require larger values of
$p$ to reproduce the observed bend in the XLF.  The primary 
effect of increasing $\Omega_0$ is to decrease $\sigma_8$ and thus
$\sigma(M)$, strengthening the exponential cutoff in the high-mass
regime and again causing a sharper bend at \Mstar; the result is the
expected increase of $p$ with $\Omega$.  The implications of this
parameter and the meaning of the limits drawn in Figure 2 are discussed
further in section 4.1. 
  
The behaviour of $L_{15}$ follows from this analysis, but is more complex.
Since the XLF has a discernable bend at about $2 \times 10^{44}$ erg 
sec$^{-1}$, the value of $\Mstar$ can be expected to have a strong influence
on the derived normalization.  Clearly a larger value of $n$ will lead to
a smaller value of $\Mstar$, since it only appears in the exponent $1/2\alpha$.
On the surface it appears that increasing $\Omega_0$ should also cause $\Mstar$
to drop, since it is proportional to $\sigma_{15}$, but this also has the
effect of raising the value of $p$.  Since 
$L_X(\Mstar) \propto L_{15}\Mstar^p$, the required value for $L_{15}$ can
be expected to display a minimum where the two effects are balanced.  If $n$
is zero, the dependence of $\Mstar$ on $\sigma_8$ is greatly weakened and
the change in $p$ dominates the behaviour of $L_{15}$.  

There are dependencies on $s$ and $\lambda_0$ as well,
but these are slight compared to those described above. Increasing $s$
decreases $L_{15}$ according to the factor $(1+z)^s$ folded into the
BCS fit, and does not affect $p$ at all. Introducing a cosmological constant
decreases $L_{15}$ and $p$ for low density models, mainly because these
universes require a larger $\delta_{c0}$ and therefore have a smaller $\Mstar$.

The ``1-$\sigma$ variations'' shown in Figure 1
are the curves obtained by using the vertices of the covariance matrix error
ellipse to modify $L_{15}$ and $p$.  The uncertainties depicted in Figure 2
are the formal 1-$\sigma$ error bars given by the diagonal elements of the
covariance matrix.  Since the functional form of the XLF is
nonlinear, these limits should be taken as a reasonable estimates of
the variance rather than precise measurements of a confidence level. 

\begin{figure}
\epsfxsize 8.5cm
\epsfysize 8.5cm
\epsfbox{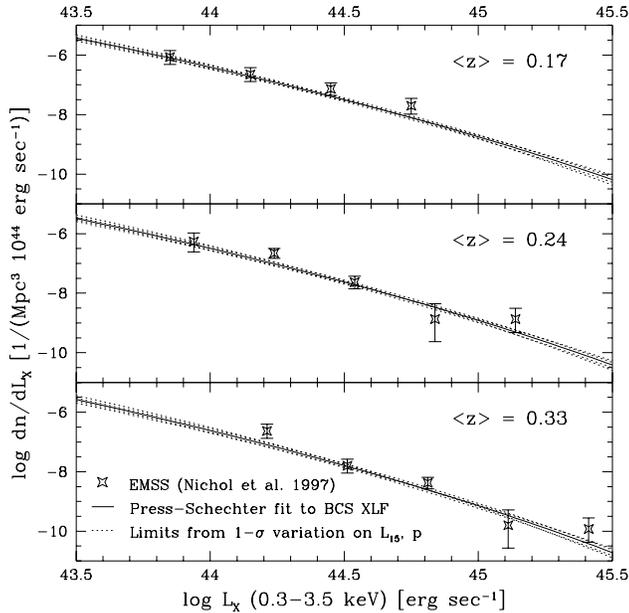}
\caption{The EMSS X-ray cluster distribution as reconstructed by Nichol \etal
(1997)  The solid line represents the same best-fit ``standard CDM'' model
that was shown in Figure 1, and the dotted lines represent 1-$\sigma$
variations.}  
\label{EMSSdat}
\end{figure}

At this point we have six parameters that describe our models: $\Omega_0$, 
$\lambda_0$, $n$, $s$, $L_{15}$, and $p$.  The first four are arbitrary, and
the last two are determined from local observations.  This is all the
information we need to construct flux-limited statistics of the
underlying Press-Schechter abundance for a given cosmology. We can make an
additional test of the method by comparing our BCS-fitted distribution to 
the EMSS cluster sample (Henry \etal 1992), recently reanalysed 
by Nichol \etal (1997), which has the advantage of being binned in redshift.
The results of this comparison are presented in Figure~\ref{EMSSdat},
along with the same variations on the theoretical XLF that were displayed
in Figure 1.  The theory with $s \se 3.5$ and data are in good
agreement, but the depth  of sample is too shallow to provide significant
leverage on $s$;  models with $s = 0.0$ or $s = 6.0$ provide equally good fits.
Putting good constraints on cluster evolution requires either a very
large sample or a very deep sample, as we will see in the next section. 

\section{Predictions and Observations}

\subsection{The logN--logS}

The model as stated above is sufficient to describe the underlying 
abundances of clusters for a variety of universes.  We wish to 
compare these abundances to recent observations of the logN--logS 
statistic.  Calculating number counts as a function of observed flux 
is done by integrating the number density of clusters over mass 
and redshift, with a lower limit on the mass integral coming from the 
minimum mass required to meet the flux level at each redshift.  

The band-limited flux $S_X(E)$ is related to the observed, band limited 
X-ray luminosity by the usual relation 
\begin{equation}
S_X(E) = \frac{L_X(E)}{4\pi r^2(1+z)^2}
\end{equation}
where $r$ is the physical distance from the observer to the cluster 
determined by the cosmological parameters within the Robertson--Walker 
metric.  For a universe with $\lambda_0 = 0$, this gives 
\begin{equation}
S_X(E) = \frac{L_{15}H_0^2q_0^4(1+z)^sM^pf_E(M)}
{4\pi c^2[q_0z+(q_0-1)(\sqrt{2q_0z+1}-1)]^2}
\label{SXvsMass}
\end{equation}
This equation allows the minimum mass satisfying a given flux limit to 
be calculated at each redshift.

Depending on the form of the observations that we are trying to imitate, we
have the option of adding in at this point a correction for the finite
cluster size and the point response function of the telescope, or some other
factor that represents the efficiency of the flux recovery.  This factor can
either be a pure number or, if you are willing to invoke a model for the
surface brightness, a function of the cluster's angular diameter
(e.g., EH91).  As the
observers have already made this correction in constructing the logN--logS,
we calculate the total flux as given above.  

The number of clusters per unit mass on the sky is 
calculated from the basic relation
\begin{equation}
dN = dn(M,z)dV = dn(M,z)\frac{r^2dr}{\sqrt{1-kr^2}}d\Omega
\end{equation}
which can, with the appropriate function $r(z)$, 
be transformed into a function of redshift
\begin{equation}
\frac{dN}{dzd\Omega} = \frac{dn(M,z)}{H_0^3(1+z)^3q_0^4}
\frac{[q_0z+(q_0-1)(\sqrt{2q_0z+1}-1)]^2}{\sqrt{1-2q_0+2q_0(1+z)}}
\end{equation}
which again holds for $\lambda_0 = 0$.  Integrating this function over 
redshift with lower mass limit given by equation~(\ref{SXvsMass}) 
gives the number of visible clusters in the sky per steradian as a function 
of limiting flux $S_X(E)$.  Similarly, 
we can construct a differential logN--logS or keep track
of the total number in each redshift bin to predict redshift distributions.

The situation for $\Omega_0+\lambda_0 = 1$ universes is slightly more
complicated in that the function $r(z)$ can not be expressed in simple 
analytic form.  It can, however, be tabulated and used to calculate the
volume element and observed flux to arbitrary accuracy. Appendix B contains
the details of this derivation.

\subsection{ROSAT Data and Comparisons}

We use logN--logS data from two independent, serindipitous 
samples of X-ray clusters derived from deep, pointed ROSAT
observations.  The work of Rosati \& Della Ceca (1997) includes 125
clusters over a total sky area of about 35 square degrees, and goes down
to a flux limit of $2 \times 10^{-14}$ ergs sec cm$^{-2}$.
The work of Jones \etal (1997) includes 34 clusters and goes
from $4 \times 10^{-14}$ to $2 \times 10^{-13}$ in $S_X$.  Both groups employ 
different methods and assumptions which we now discuss.  

Rosati \& Della Ceca (1997) use a wavelet decomposition
algorithm to identify extended 
sources in ROSAT PSPC fields, and employ the wavelet coefficients to
reconstruct the total flux of each source.  This technique is subject to large
random errors on clusters of low signal-to-noise ratio,
but is no worse off in these terms than fitting the cluster profile to a
standard (e.g., King) model.  There is also a bias
inherent in reconstructing the area under a non-gaussian profile with a
gaussian wavelet, but this is small compared to the sources of random error in
the problem.  In his doctoral thesis, Rosati (1995) presents 
details of the reconstruction process.  He also implements a sophisticated
correction for the sky coverage, which takes into account distortion in 
PSPC images at large off-axis angles and varies with the angular diameter and
flux of the source.  Uncertainties in the integral counts
include Poisson noise, uncertainties in the
sky coverage fraction, and the random errors inherent in reconstructing the
flux of an image with low signal-to-noise ratio; the error bars used in
the differential counts, however, are just Poisson.

The work of Jones \etal is derived from the WARPS cluster survey of 
Scharf \etal (1997), which uses the VTP analysis method (Ebeling \&
Wiedenmann 1993; Ebeling 1993) to detect sources.  To account for
any flux that might be hidden under the background level, they assume
a standard $\beta$-model form with $\beta = 2/3$, derive a
normalization and core radius from the source profile, and integrate
in the region of low signal-to-noise ratio
to find an appropriate correction factor.  The model is not
used to determine the overall flux, just to make a second-order
correction to the VTP count rates.  Details of their flux 
correction procedure can be found in Ebeling \etal
(1996). The error bars of this dataset are strictly Poisson and are claimed to 
dominate all other sources of error.  Numbers used in this analysis were 
kindly made available by L. Jones prior to publication.

Both groups attempt to construct the logN--logS in a largely 
model independent fashion.  The advantage of this approach is that it 
avoids building in biases based on questionable assumptions.  The 
disadvantage is that it makes it difficult to assess sample completeness 
in a systematic fashion; it is unclear how large a population may be 
missing from the detected counts because of low surface brightness and/or
large angular extent.  Of course, comparing results of the independent 
groups is a simple gauge of systematic effects.  The data, shown in 
Figure 4, indicate consistency within modest statistical errors.  
Given that it is easier to err in the direction of missing extended,
low surface brightness sources, it is reasonable to interpret the
current data as providing firm lower limits to the true cluster counts,
and accurate estimates if no such population exists.  Thus, a model
which {\em underpredicts} the number of clusters can be excluded with
somewhat greater confidence than one which overpredicts the observed counts.
 
\begin{figure}
\epsfxsize 8.5cm
\epsfysize 8.5cm
\epsfbox{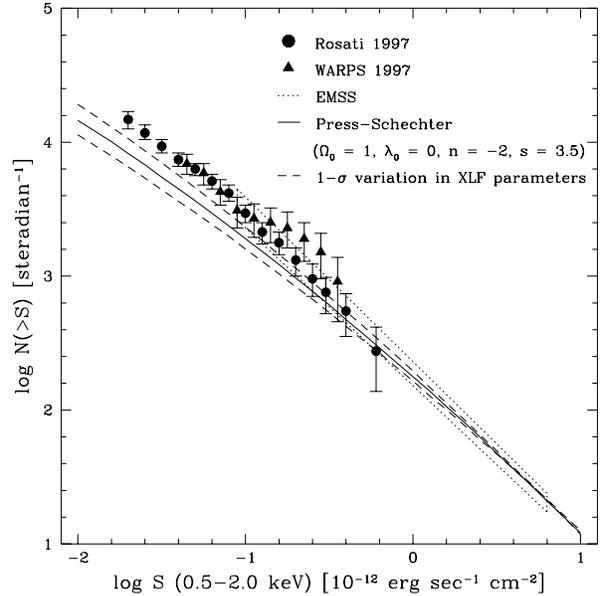}
\caption{The integrated cluster surface density above an X-ray flux limit $S$.
The solid line respresents the best-fit values of $L_{15} \se 3.556$ and 
$p \se 2.174$, while the dashed lines represent 1-$\sigma$ variations around
these values as described in the text and displayed in Figures 1 and 3.}  
\label{lnls}
\end{figure}

\begin{figure}
\epsfxsize 8.5cm
\epsfysize 8.5cm
\epsfbox{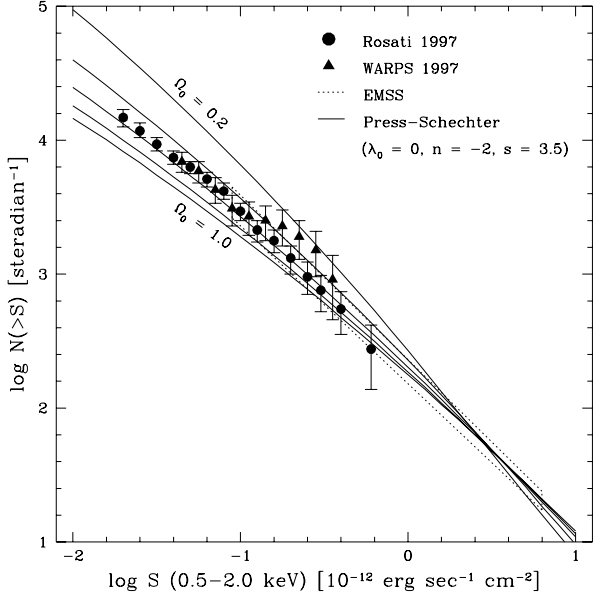}
\caption{The integrated cluster surface densities for models with 
$\lambda_0 \se 0$, $n \se -2$, and $s \se 3.5$, while 
$\Omega_0$ takes on the values 0.2, 0.4, 0.6, 0.8, and 1.0}  
\label{omlnls1}
\end{figure}

\begin{figure}
\epsfxsize 8.5cm
\epsfysize 8.5cm
\epsfbox{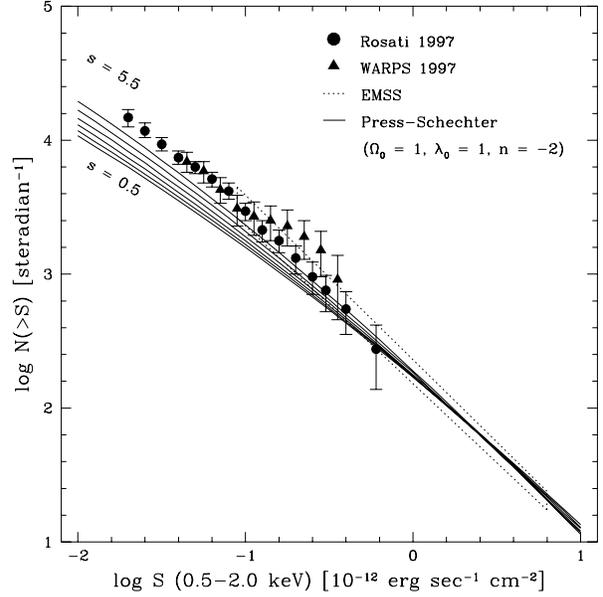}
\caption{The integrated cluster surface densities for models with 
$\lambda_0 \se 0$, $\Omega_0 \se 1.0$, and $n \se -2$, while 
$s$ takes on the values 0.5, 1.5, 2.5, 3.5, 4.5, and 5.5.}  
\label{slnls}
\end{figure}

\begin{figure}
\epsfxsize 8.5cm
\epsfysize 8.5cm
\epsfbox{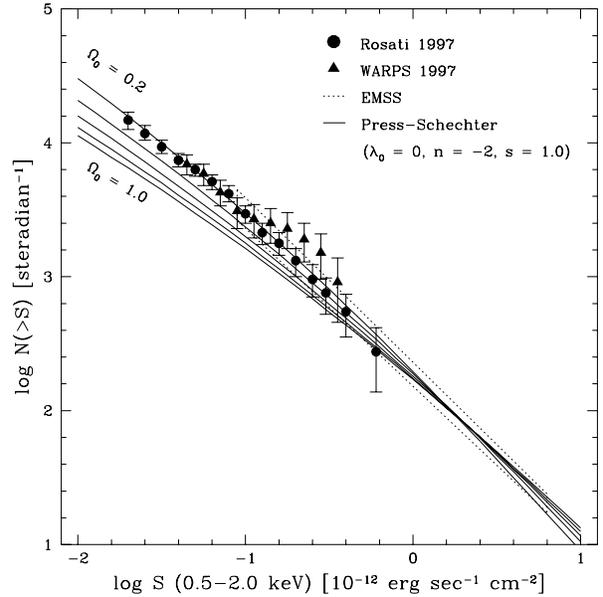}
\caption{The integrated cluster surface densities for models with 
$\lambda_0 \se 0$, $n \se -2$, and $s \se 1.0$, while 
$\Omega_0$ takes on the values 1.0, 0.8, 0.6, 0.4, and 0.2.  Note that a
lower value of $s$ greatly supresses the differences between low-density
and high-density models, because high-redshift clusters become harder to
observe.}  
\label{omlnls2}
\end{figure}

Figure~\ref{lnls}
gives an example of the integrated logN--logS for a  
universe with $\Omega_0 = 1$, $\lambda_0 = 0$, $s = 3.5$, and $n = -2$.
Superimposed are the observational data from Rosati \& Della Ceca (1997) 
and the WARPS sample.  The data are shown in their familiar cumulative form
here, so the points of each sample are not statistically independent from
one another.  A crude judgement of goodness 
of fit can be made by comparing the model to the faintest points of 
each sample.  This particular model underpredicts the number of clusters, 
to a moderate but still significant degree.  This is true 
even for the logN--logS curves that arise from modifying $L_{15}$ and $p$
through the vertices of their one sigma error ellipse, shown as the 
dashed lines in the figure.  

Figures 5 through 7 demonstrate the sensitivity of the counts to variations
in $\Omega_0$ and $s$.  Figure~\ref{omlnls1} displays the changes that occur as
$\Omega_0$ decreases from 1.0 to 0.2 while $s$ is held constant at a value of 
$3.5$.  Clearly, low $\Omega$ universes produce higher counts, a reflection 
of the slower evolution experienced by the cluster population in such 
models (Richstone, Loeb \& Turner 1992) as well as the larger 
volume element per redshift interval.  Figure~\ref{omlnls2}
displays the logN--logS for the same five universes but with a mild evolution
parameter $s \se 1.0$.  Reducing $s$ has the effect of reducing
the differences between universes with different density parameters as well
as lowering the overall number of cluster, since a lack of positive luminosity
evolution quickly dims the high-redshift population expected in open models.
This extinction has a proportionally larger effect on the low $\Omega_0$ 
models, for which a given flux limit represents a deeper probe.  

The effect of changing $s$ while holding $\Omega_0$ fixed is shown in
Figure~\ref{slnls}.  
The variation in counts seen here is a combination of two factors: first, a 
model with less positive evolution in the cluster luminosity will contribute
fewer objects to the logN--logS at a given flux limit; and second, the overall
normalization of the mass--luminosity relation will be larger if $s$ is
smaller.  (The second effect arises from the approximation that the entire
BCS sample lies at an average redshift of 0.1, which introduces a dependence
of $L_{15}$ and $p$ on $s$--see equation~(\ref{LMz_relation}).)  
The two effects change in strength and work in opposite directions, 
so the variation shown in Figure 6 cannot be taken as universal.  
In particular, low-density models display a much stronger dependence on
$s$ than models in which $\Omega_0 \se 1$; this can be understood as an
increase in the importance of the high-redshift population relative to
the local sample.

\subsection{ Examination of the $\Omega_0 - s$ plane }
% All three of these figures should rest on the same page if at all possible

\begin{figure}
\epsfxsize 8.5cm
\epsfysize 8.5cm
\epsfbox{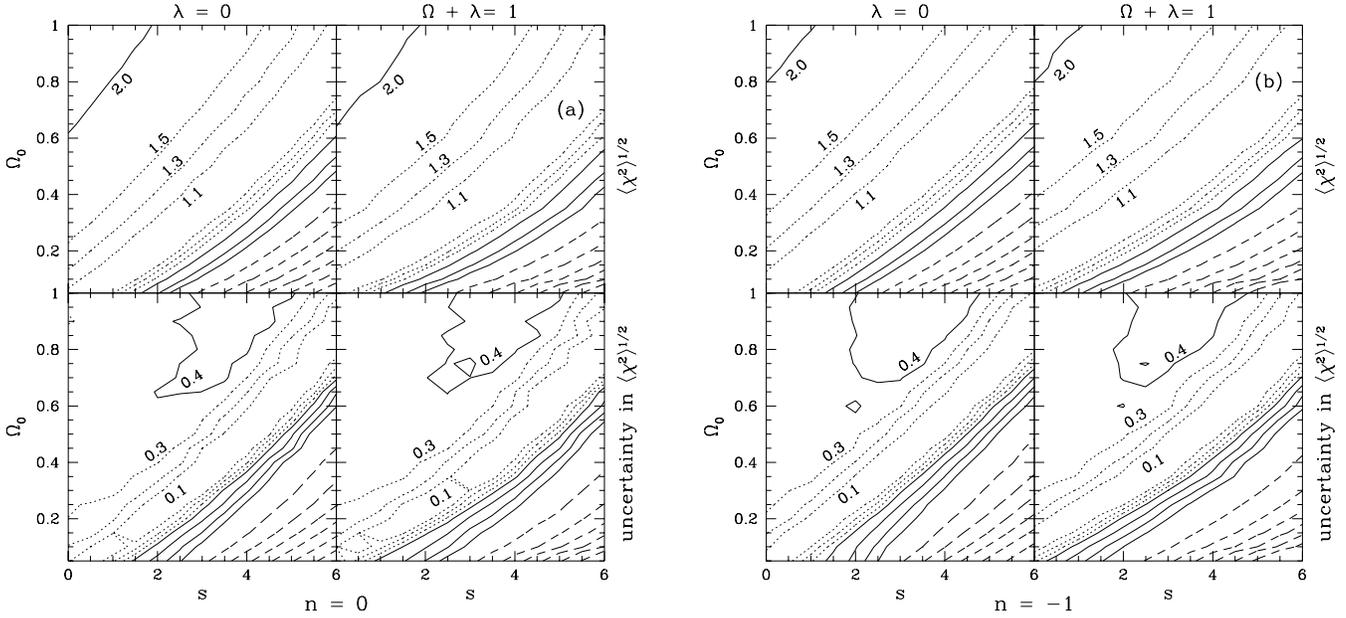}
\caption{The square root of the reduced chi-squared parameter for all models 
is displayed, as well as the ``uncertainty'' in this parameter, the 
derivation of which is described in the text.  In the upper row, dotted-line
contours have been drawn corresponding to $\langle\chi^2\rangle^{1/2}$ = 1.07,
1.28, and 1.51, which represent formal probabilities of about 30\%, 5\%, and
0.3\% that the data come from the same underlying distribution as the model
and differ only through gaussian random variations.  Solid-line contours
have been drawn at $\langle\chi^2\rangle^{1/2}$ = 2.0, 3.0, and 4.0, which
correspond to very small probabilities but may be useful in combination with
the uncertainties.  The dashed lines are drawn every time the likelihood
parameter doubles thereafter, {\em i.e.} at levels of 8.0, 16.0, etc.
In the lower row, the dotted-line contours are drawn at uncertainty levels
of 0.1, 0.2, and 0.3; the solid lines at levels of 0.4, 0.6, 0.8, and 1.0, 
and the dashed lines at each doubling over 1.0. Note that although the
uncertainties get very high in the region of low $\Omega_0$ and high $s$,
the likelihood parameter is larger still by about a factor of 4.
Figure (8a) displays models with $n \se 0$, Figure (8b) with $n \se -1$
and Figure (8c) with $n \se -2$.}
\label{omegas}
\end{figure}

\begin{figure}
\epsfxsize 8.5cm
\epsfysize 8.5cm
\epsfbox{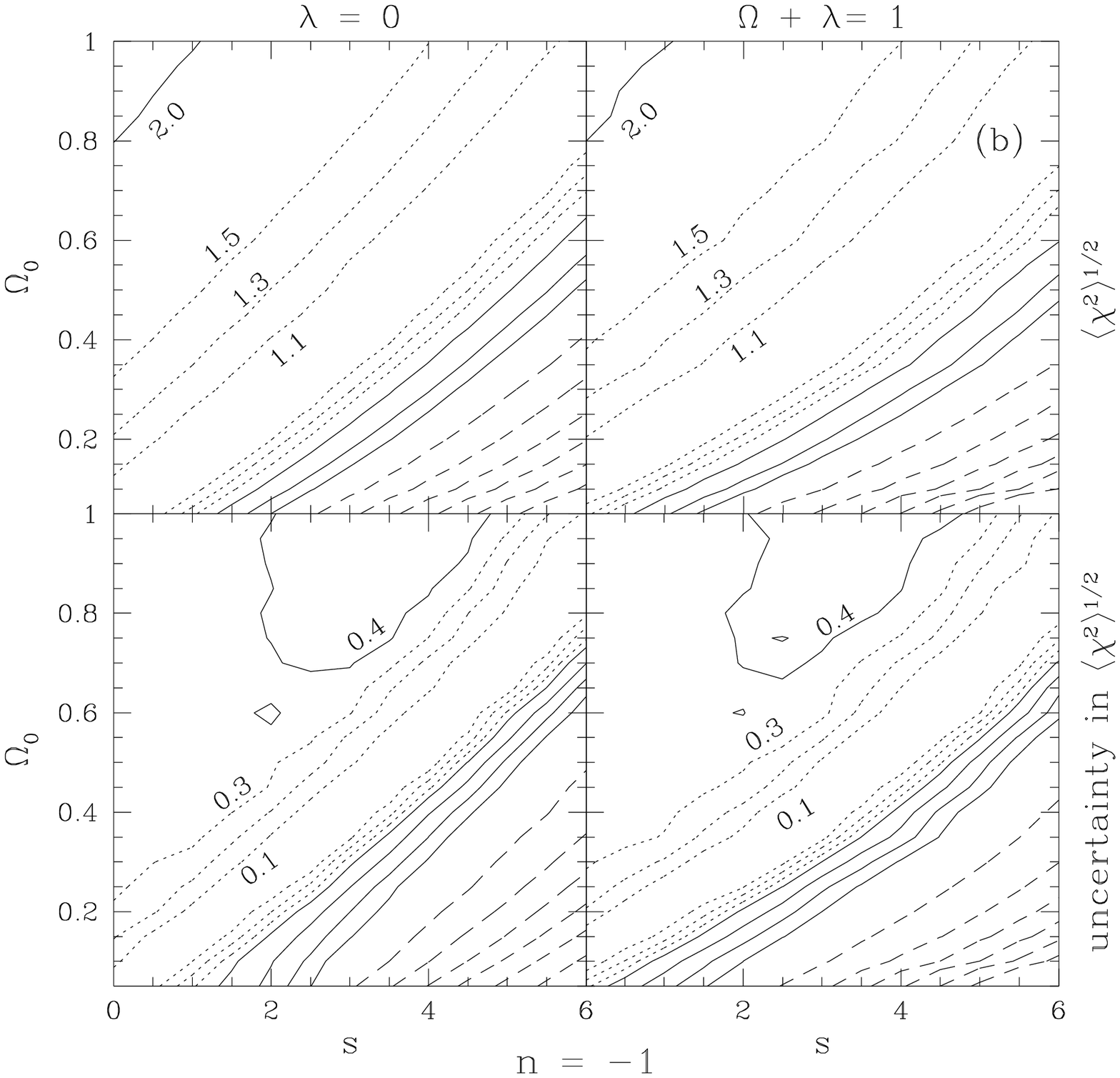}
\end{figure}

\begin{figure}
\epsfxsize 8.5cm
\epsfysize 8.5cm
\epsfbox{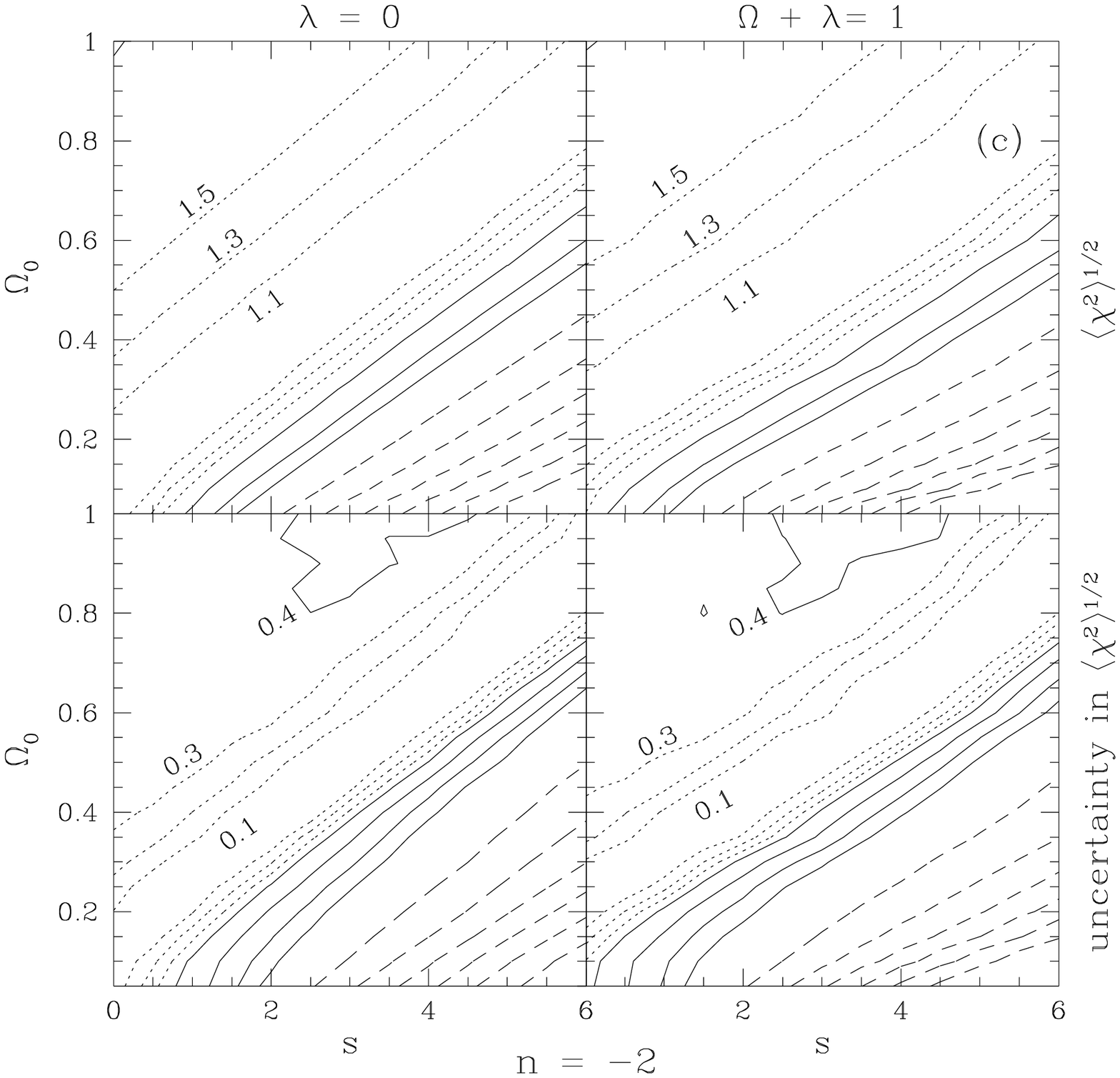}
\end{figure}

Changing the cosmological parameters over the proposed range can
produce massive overabundances or moderate underabundances.  To assess the
likelihood of each model we compare the predicted differential logN--logS
(the number of clusters in a specified flux bin) to the data points and 
calculate the reduced chi-squared factor for that model.  The full set of
data used to constrain our parameter space includes the faintest 6 data points
from the WARPS sample and ten points from Rosati's,
for sixteen degrees of freedom altogether.  The use of 
differential data means that statistical errors are formally independent.  
We employ a chi-squared statistic because it provides a measure of the 
absolute goodness of fit for the models, assuming the error bars 
are Gaussian.  Our results would be similar 
were we to employ a maximum likelihood approach.  The main 
disadvantage of this approach is the difficulty in including the 
uncertainty in $L_{15}$ and $p$ for a particular model.  
We adopt an approximate procedure to solve this problem, which will be
described shortly.

Figures~8(a-c) present the 
``likelihood'' plots of all models having $n = 0$, $-1$ and $-2$, respectively.
The value displayed in the upper row for each model is
the square root of the reduced chi-squared parameter 
$\langle\chi^2\rangle^{1/2}$, where the average is over all data points
in the logN--logS.  
The dotted contours in the upper row mark models which deviate from the data
at one, two, and three-sigma confidence levels.
The main limitation in the ability of theory to predict abundances accurately
at this level of analysis is the determination of $L_{15}$ and $p$.  Although
an accuracy of 5\% to 10\% is quite good (and quite possibly as good as we're
going to get, since this method is limited by the number of nearby clusters),
varying these numbers within their confidence limits can still produce
considerable changes in the predicted logN--logS (see Figure 4).

To account for this, we calculated the likelihood parameter, 
$\langle\chi^2\rangle^{1/2}$, for the four vertices of the
standard error ellipse in $L_{15}$--$p$
space as well as for the best-fit model.  The uncertainty shown is the
difference in $\langle\chi^2\rangle^{1/2}$ between the best-fit case and the
perturbed case which gave the smallest value of the likelihood parameter.
The results are presented in the lower rows
of Figures~8(a-c) 
and can be interpreted as a ``one-sigma'' uncertainty in the
contours of the top row.  Keep in mind that this variance is not symmetric; 
we present only the difference in the direction of greater likelihood.
Since our aim here is to provide a reasonable indication of the uncertainty 
in the modeling, this heuristic error estimate seems sufficient.  The data
were presented in this manner to make it possible for the reader to
estimate constraints at other confidence levels. 

It is clear from Figures 8(a-c) that the results are not strongly
affected by the value of $\lambda_0$.
Introducing a cosmological constant increases the volume element per
redshift bin throughout most of the integration range, which reduces
the observed flux of each cluster, but it also increases the total surface
density of clusters.  As can be seen from a close examination of the
figures, in most regions of the parameter
space introducing a cosmological constant slightly 
raises the value of $\Omega_0$ required for a good fit.  Since raising
$\Omega_0$ lowers the number of clusters, we conclude that
increased integration volume is the dominant effect for most models.
In other words, setting $\lambda_0 \neq 0$ tends to increase the surface
density of clusters, and a higher value of $\Omega_0$ is needed to
compensate.

Changing the spectral index $n$ has a quite strong
effect on the derived mass-luminosity relationship (Figure~2), which
as we will see can have important consequences.
Because all of the models are constrained to agree with local abundances, 
however, $n$ has limited leverage on the distant cluster population.
The main effect on the logN--logS is that models with larger $n$ have
sharper exponential cutoffs in their mass distribution, resulting in a
lower cluster surface density.  The best-fitting models for $n \se 0$ thus
have lower $\Omega_0$ and greater luminosity evolution to compensate.  

The clearest relationship visible in Figures~\ref{omegas}(a-c) 
is, of course, that
between the density and evolution parameters, $\Omega_0$ and $s$. 
It seems from these results that if we wish to believe in the degree of
evolution indicated by self--similar or constant entropy arguments ---
$s \se 3.5$ and $11/4$, respectively ---  then we obtain a useful 
lower bound of $\Omega_0 \gta 0.3$.   The lower end of this range is 
consistent with limits from the mean intracluster gas fraction of 
clusters, which give $\Omega_0 h^{2/3} \se 0.28 \pm 0.07$ (Evrard
1997) in the most straightforward interpretation.  

Conversely, if the degree of luminosity 
evolution is instead very small --- $s \lta 1$ --- then the analysis 
strongly rules out a critical density universe.  Recall 
that all the off-ridge models which have 
$\Omega_0 \se 1$ underpredict the number of observed clusters 
(see Figures 5-7), and these should be treated more harshly given
the possibility for survey incompleteness.
It seems that if the universe does indeed have a critical density, we need
at least a moderate degree of evolution in the cluster luminosity to
fit the data well.  The assumption of constant cluster entropy ($s = 11/4$)
put forward in EH91 would be sufficient to allow such models.
The preferred region, however, indicates an even larger degree of
evolution.  This could come about if radiative cooling occurs on a much longer
time scale than gravitational collapse (Bower 1997), or 
if a significant amount of energy
is injected into the ICM by galaxies (e.g., Metzler \& Evrard 1994).

\begin{table*}
\begin{tabular}{lcccccc} \hline
& \multicolumn{3}{c}{$\lambda_0 = 0$} &
\multicolumn{3}{c}{$\Omega_0 + \lambda_0 = 1$} \\
Constraint & $n = 0$ & $n = -1$ & $n = -2$ & $n = 0$ & $n = -1$ & $n = -2$
\\ \hline
$\Omega_0 = 1.0$ & $s \geq 3.5$ & $s > 3.0$ & $s > 2.5$
& $s \geq 3.5$ & $s > 3.0$ & $s > 2.5$ \\
$s = 3.5$ & $\Omega_0 > 0.20$ & $\Omega_0 > 0.25$ & $\Omega_0 > 0.35$
& $\Omega_0 > 0.20$ & $\Omega_0 > 0.25$ & $\Omega_0 > 0.35$ \\
$s = 1.0$ & $\Omega_0 \leq 0.45$ & $\Omega_0 < 0.60$ & $\Omega_0 \leq 0.75$
& $\Omega_0 \leq 0.45$ & $\Omega_0 < 0.65$ & $\Omega_0 < 0.80$ \\
$s = 0.0$ & $\Omega_0 < 0.30$ & $\Omega_0 \leq 0.40$ & $\Omega_0 \leq 0.60$
& $\Omega_0 < 0.35$ & $\Omega_0 \leq 0.45$ & $\Omega_0 \leq 0.65$ \\ \hline
\end{tabular}
\caption{A summary of the constraints that can be placed on $\Omega_0$ and
$s$ using the information in Figures~8.  All of these constraints are at
the 95\% confidence level, after accounting for the displayed uncertainty
in the chi-squared parameter.}
\label{constraints}
\end{table*}

The results of this analysis are summed up in Table~\ref{constraints}.
The limits put down are at the 95\% (about 2 $\sigma$) confidence level
(i. e. there is only a 5\% chance of the model randomly producing the
observed data), {\em after} reducing the $\langle\chi^2\rangle^{1/2}$
parameter by the uncertainty indicated in Figures 8(a-c).  The models are
examined in intervals of 0.05 in $\Omega_0$ and 0.5 in $s$, so a $\geq$ or
$\leq$ sign just means that the limiting model was at or very near the
required confidence level of 95\%.  In addition, all of the disallowed models
indicated here underpredicted the number of clusters, except for those
which assumed $s \se 3.5$. The constraints thus allow
for a one-sigma variation in $L_{15}$ and $p$ tailored to bring the
model closer to the data, and can be treated as conservative conclusions.

\section{Additional Constraints from the Cluster Population}

\subsection{The Luminosity-Temperature Relation}

Because of the degeneracy between intrinsic luminosity evolution and 
cosmological evolution, the logN--logS alone limits models to the
ridge seen in Figure~\ref{omegas}.  
It is worthwhile to see what additional constraints can be
placed on the parameter space by including additional observational
information.  One interesting question is what each model predicts for the
\xray luminosity---temperature correlation.  So far, we have made
use of the virial relationship $T \propto M^{2/3}$ only to 
tabulate the band fraction.  
Because the fraction $f_E(T)$ for the ROSAT bands is a weakly 
dependent function of temperature, our results up to this point would change
very little if we were to adopt a different $T(M)$ behavior.  If we
are willing to promote virial equilibrium to a strong assumption, however,
we can use it as an independent test of the parameter $p$.  
There is good theoretical support from numerical simulations for
virial equilibrium within the non--linear portions of clusters
(Evrard \etal 1996) and modest empirical support for
this assumption from analysis of the mean intracluster gas fraction
(Evrard 1997).  Comparison with the values derived from the BCS XLF 
fit then provides an added, non--trivial constraint 
on the models.  

The virial scaling assumption $T \propto M^{2/3} (1+z)$ translates
directly into a bolometric $L_X$--$T$ relation of the form 
\begin{equation}
L_X  \ \propto \ T^{3p/2}(1+z)^{s-3p/2}
\label{LxT}
\end{equation}
The observed relationship is, roughly, $L_X \propto T^{2.8\pm0.3}$.
(Edge \& Stewart 1991; Arnaud 1994).  
The quoted error in the slope is more generous than those reported
in individual works, in order to allow for possible systematic uncertainties
betwen different data sets.  Consistency with the observed slope 
requires $p = 1.9 \pm 0.2$.  This range, drawn in Figure 2 as dotted
horizontal lines, incorporates the value $11/6$ 
appropriate for the constant central entropy model of EH91 (see also
Bower 1997). A quick glance at the figure reveals that including this
result puts a strong constraint on $\Omega_0$ and $n$. If $n = -1$, this
range of values is found only in very low-density universes
($\Omega_0 \lesssim 0.1$). For models with $n = -2$, on the other hand,
we find acceptable values of $p$ in the range $0.25 \leq \Omega_0 \leq 1.0$.
If a cosmological constant is included, the allowed range in 
$\Omega_0$ shifts to slightly higher values.  

The real perturbation spectrum is not likely a pure power law, but
this analysis indicates models with effective spectral index 
$n \lesssim -1.5$ are favored in universes with reasonable 
values of the density parameter.  Such ``red'' spectral values 
are also favored by the shape of the temperature abundance function 
(Henry \& Arnaud 1991; Oukbir, Bartlett \& Blanchard 1997).  
For reference, Huss, Jain \& Steinmetz (1997) present a list of
effective spectral indices ($n_{eff} \sequiv \dln P(k) / \dln k$) 
for a number of popular cosmological models.  A cold, dark matter
model with $\Omega_0 \se 0.3$, $\Lambda \se 0$ and $h \se 0.7$ has 
$n_{eff} \se -1.74$, for example. 

Room to maneuver around this conclusion can be gained by invoking
a degree of scatter in the relationship between bolometric X-ray luminosity
and binding mass.  Scatter will tend to flatten the slope and increase
abundance at the bright
end of the predicted luminosity function, implying a bias toward 
overestimating $p$ and $L_{15}$ compared to the case with no scatter.  
The amount of scatter is not known {\sl a priori\/}, but a likely 
lower bound can be obtained from gas dynamic simulations.  
The two sets of 18 cluster simulations of Metzler (1995) revealed  
a scatter of magnitude $\langle\delta(\log L_X)^2\rangle = 0.047$ for
simulations involving dark matter and baryon fluids, and
$\langle\delta(\log L_X)^2\rangle = 0.026$ for simulations which also
included a model for the ejection of gas from galaxies into the ICM. 
We have checked that including scatter at the larger of the above 
values results in a fractional 
change in the best fit $p$ of at most $10\%$, comparable to the
uncertainty from the BCS XLF fit.  We conclude that such modest levels
of scatter are not strongly affecting our results.  Larger
amounts of scatter would lead to downward revisions in our best fit
values of $p$ in Figure~2, and so would somewhat relax constraints on
the effective spectral index on cluster scales.  

Another important prediction from equation~(\ref{LxT}) is that we expect 
redshift evolution in the bolometric luminosity at a fixed temperature
proportional to $(1+z)^{s-3p/2}$.  Since fitting the local relation
requires $3p/2 \se 2.8 \pm 0.3$, the expectation is of an offset 
in the intercept of the $L_X - T$ relation of amplitude 
\begin{equation}
\Delta(\log L_X(z)) \, {|}_T \ = \ [s-(2.8 \pm 0.3)] \, \Delta(\log(1+z)) 
\label{LxTshift}
\end{equation}
independent of spectral index or cosmology.  Both the original 
self--similar value $s \se 7/2$ and the constant central entropy value 
$s = 11/4$ predict very little change in the intercept of the 
\LxT relation back to $z \ssimeq 0.5$.  The full range we investigate,
$s \in [0,6]$, translates into shifts of factors ranging from about
one--third to three at $z=0.5$.
   
Comparing a recent compilation of 15 distant clusters
($\langle z \rangle \simeq 0.33$) by Henry \etal (1997) with a similar
sample of the nearby population ($\langle z \rangle \simeq 0.07$) 
by David \etal (1993) provides us with an early estimate of this shift.
Comparing the intercepts of the two samples, we find
$\Delta(\log L_X)/\Delta(\log(1+z)) \se 1.1 \pm 1.1$
at the one-sigma level, which corresponds to a plausible range of
$2.5 \leq s \leq 5.3$.  The observations rule out very low
values of the evolutionary parameter $s$, and therefore eliminate the
low $\Omega_0$ end of the allowed ridge in Figure~\ref{omegas}.

\subsection{Redshift Distributions}

\begin{figure}
\epsfxsize 8.5cm
\epsfysize 8.5cm
\epsfbox{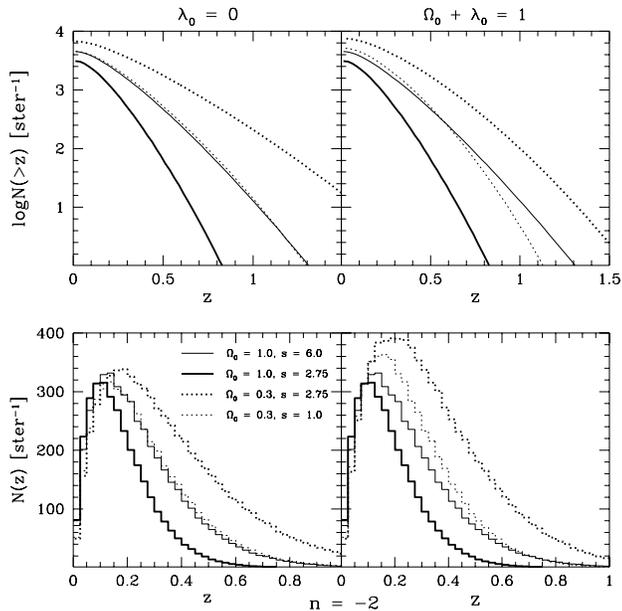}
\caption{The cumulative and differential redshift distributions for the
$\lambda \se 0$, $n \se -2$ models described in the text.  The effect
of changing $\Omega_0$ and $s$ can be clearly seen here; increasing 
$\Omega_0$ sharply reduces the number of high-redshift clusters, while
increasing $s$ brings the lower-mass objects above the flux limit.}
\label{redshifts}
\end{figure}

Additional constraints can be found by examining the redshift
distribution of the clusters, which can be expected to vary strongly
with our choice of cosmology.  Indeed, 
Oukbir \& Blanchard (1997), employing methods similar to ours, found
the redshift distribution in the EMSS sample (Gioia \& Luppino 1994) 
to be well fit by either a low density model with mild evolution
($\Omega_0 \se 0.2$, $s \se 0.5$ in our terminology) or an
Einstein--deSitter model with more rapid evoluition ($\Omega_0 \se 1$,
$s \se 3.8$).  These models lie along the ridge allowed by the ROSAT
counts in our analysis.  

As discussed in section 3.1, the redshift
distribution is straightforward to calculate for a given model and flux limit.
Although obtaining accurate redshifts for clusters is a time-consuming process,
the shape of this statistic is not strongly dependent on small
differences in the efficiency of flux reconstruction or the choice of
model for cluster surface brightness. A confident flux limit and a
well-understood sky coverage function, are, however,
necessary to predict the normalization of these curves.

Figure~\ref{redshifts} displays the differential and cumulative distributions
at a flux limit of $5.5 \times 10^{-14}$ erg sec$^{-1}$ cm$^{-2}$ in the band
0.5-2.0 keV, the parameters of the WARPS survey.  Each category of models
is represented by four specific choices within the region of plausible
likelihood, all of which assume $n \se -2$:

\begin{enumerate}
\item $\Omega_0 = 0.3$, $s = 1.0$
\item $\Omega_0 = 0.3$, $s = 2.75$
\item $\Omega_0 = 1.0$, $s = 2.75$
\item $\Omega_0 = 1.0$, $s = 6.0$
\end{enumerate}

Models (i) and (iv) fit all of the logN--logS data quite well, lying near the
center of the ridge described by Figure 8c, but model (i) predicts a
decrease in the zero--point of the $L_x$--$T$ relation larger than observed.  
Models (ii) and (iii) are given
as reference points, and to display the effects of moving perpendicular
to the ridge in the $\Omega_0$--$s$ plane. These models are ruled out at about
the 95\% confidence level after subtracting the uncertainty in their
likelihood parameters, but make the effect of changing the parameters clearer.

As you can see, the redshift distribution does not discriminate well between
models which lie on the ridge in Figure 8c, but displays significant
differences when the parameters are changed in other directions.
Samples of size approaching 100 clusters are soon arriving (Rosati
1997, private communication) and these will provide stringent tests of
the models by constraining the 
area under the high redshift tail of the distribution.  
In particular, if $\Omega_0 \simeq 0.3$ as indicated by the mean 
intracluster gas fraction and $s$ is close to the $11/4$ fixed core
entropy value (model ii), then we predict 10\% of clusters above the 
WARPS survey flux limit should lie at $z \ge 0.75$.  

\section{Conclusions and Future Directions}

We assume that the underlying number density of clusters 
as a function of mass is well described by the Press-Schechter 
function and model the scaling of cluster bolometric luminosity 
and temperature as power--laws in binding mass and epoch.  
For a given cosmological model, we determine the
free parameters of the current mass-luminosity relation by 
fitting the predicted \xray luminosity function 
to the BCS XLF (Ebeling et al. 1996), and integrate the abundance 
function out to high redshifts to compute counts.  

A direct comparison by means of chi-squared analysis made to differential
logN--logS data from two ROSAT surveys produces a ridge of acceptable 
models following roughly $s \simeq 6 \Omega_0$.  
If $s$ is in fact about equal to 3 as expected from
simple theoretical arguments (Kaiser 1986, EH91) then the universe seems to
require at least a moderate density parameter $\Omega_0 \gta 0.3$ 
in order to bring the cluster
number counts down to observed levels.  Conversely, a universe with critical
density is inconsistent with a low evolution parameter $s \lta 2.5$;  
such models
are strongly ruled out because they underpredict the number of clusters
and are harder to save by invoking systematic errors in the surveys.
Although changing the slope of the power spectrum and incorporating a
cosmological constant also affected these results, these parameters
contribute only slightly to the overall behaviour.  An independent
constraint from the local, observed $L_X$--$T$ relation 
limits singles out the $n \se -2$ plane, further reducing the space of
allowed models.

The logN--logS alone is a useful, but limited, discriminator of 
cosmological models.  Because the local XLF is used to determine the
mass-luminosity relationship, the predicted counts will always be 
limited in accuracy by the level of shot noise in the local cluster 
population.  Assuming that the
local XLF will not increase much in accuracy, this method has the potential to
pin down a region in the $\Omega_0$-$s$ plane with a width of about $\pm 0.1$
in $\Omega_0$ and $\pm 1.0$ in $s$.  This estimate was made under the 
assumption that the uncertainties arising from the XLF are the only
significant ones; in practice this level of discriminating power could be
achieved when the overall uncertainties on the logN--logS data
have decreased to about one-third their current size.  

One could try to improve the accuracy of this method by
pushing observations down to a lower flux limit, but the returns are 
limited by the local XLF uncertainty. In order for a new data
point at a fainter flux limit to be useful, it needs to have error bars
significantly smaller than the uncertainty in the logN--logS from the
local XLF.   For example, the uncertainty on the integrated logN--logS at
a flux limit of $10^{-15} \ergs$ is about 30\%.  We estimate that a data point
with 10\% error bars would make a useful contribution to the
discriminating power of this method.  Assuming simple Poisson errors, this
level of precision could be achieved with a sample of about 100 clusters,
over 1-2 square degrees of sky. Toy models show that such a point would
strengthen the lower limit on $s$ for critical density universes, and
greatly increase the lower limits on $\Omega_0$ for models with moderate $s$.

Independent observations can be used to narrow the allowed range 
in parameter space.  Assuming clusters are in virial equilibrium, 
the shift in the zero--point of the $L_x$--$T$ relation limits the
evolutionary parameter $s$.  Current observations at
$\langle z \rangle \sim 0.3$ (Henry \etal 1997; Mushotzky \& Scharf 1997) 
indicate $s$ lies in the range $2.5$ to $5$.  Similar data at higher
redshifts, with subsequently longer lever arm in 
equation~(\ref{LxTshift}), would be very useful in reducing the allowed
range.  Redshift distributions of \xray flux limited samples provide
additional independent constraints (Oukbir \& Blanchard 1997; Bower
1997).  The extent of the high redshift tail of the
distribution is a sensitive indicator of which side of the
$\Omega_0$--$s$ ridge the universe lies.  
If $\Omega_0 \simeq 0.3$ as indicated by the mean 
intracluster gas fraction (Evrard 1997) and $s \ge 11/4$, 
then we predict at least 10\% of clusters above the 
WARPS survey flux limit should lie at redshifts $z \ge 0.75$.  

\section*{Acknowledgements}
We received much help and support from other researchers in creating this
paper;  our deepest appreciation goes out to L. David, H. Ebeling, L. Jones,
B. Nichols, P. Rosati, C. Scharf, and A. Vikhlinin for their generous
sharing of data and willingness to discuss details of their techniques.
This work was supported in part by NASA through Grant NAG5-2790, and by the
CIES and CNRS at the Institut d'Astrophysique in Paris.  AE is 
grateful to S. White for support during a visit to MPA--Garching where
much of this paper was written.

\appendix

\section[]{Calculating $\delta_c(z)$} 

The critical overdensity $\delta_c(z)$ is defined as the linearly extrapolated
overdensity of a perturbation that has just collapsed at redshift $z$.  For
universes with $\Omega_0 = 1$, this number is the canonical $\delta_c = 1.686$;
for open universes or flat universes with $\lambda_0 \ne 0$ this overdensity
will be somewhat smaller.  Since the CDM power spectrum used in our analysis is
normalized to the COBE microwave background measurements and therefore the
power spectrum at the present day, the critical overdensity of a
perturbation collapsing at redshift $z$ is further extrapolated to a redshift
of zero using the linear growth factor appropriate to the cosmology. This
quantity is labeled $\delta_{c0}(z)$.  All the following relations are derived
using the spherical collapse model.

For a flat universe with $\Omega_0 = 1$, we use the following relations
\begin{eqnarray}
\delta_c(z) & = & \frac{3}{20}(12\pi)^{2/3} \\
\delta_{c0}(z) & = & (1+z)\delta_c(z)
\end{eqnarray}

For an open universe with $\Omega_0 < 1$ and $\lambda_0 = 0$ we use the
relationship derived by Cole and Lacey (1993)
\begin{eqnarray}
\delta_c(z) & = & \frac{3}{2}D(z)\left(1+\frac{2\pi}{\sinh(\eta)-\eta}\right)
^{2/3} \\
\delta_{c0}(z) & = & \frac{D(0)}{D(z)}\delta_c(z) \\
D(0) & = & 1 + \frac{3}{x_0} + \frac{3\sqrt{1+x_0}}
{x_0^{3/2}}\ln(\sqrt{1+x_0}-\sqrt{x_0})
\end{eqnarray}
where $x_0 \equiv \Omega_0^{-1}-1$, $\eta \equiv \cosh^{-1}(2/\Omega(z) - 1)$,
and $D$ represents the linear growth factor.

Finally, for a flat universe with $\Omega_0+\lambda_0=1$ we use an approximate
parameterization to the Eke et al. (1996) results for $\delta_c(z)$
\begin{equation}
\delta_c(z) = 1.68660[1+0.01256\log\Omega(z)]
\end{equation}
The form of the parameterization was inspired by Kitayama and 
Suto (1997). This
reference also contains the functional form of the exact solution, which is a
hypergeometric function of type (2,1).  To get
$\delta_{c0}(z)$ for this type of universe we use the solution for the linear
growth factor found in Peebles (1980), \S 13
\begin{equation}
D_1(x)  =  \frac{\sqrt{(x^3+2)}}{x^{3/2}}\int_0^x x^{3/2}_1(x_1^3+2)^{-3/2}dx_1
\end{equation}
where $x = a/a_e$, and $a_e = [(1-\lambda_0)/(2\lambda_0)]^{1/3}$, the
inflection point in the scale factor.  In our analysis this function is
integrated numerically to find the growth factor at redshifts $z$ and 0,
and then $\delta_{c0}(z)$ again equals $\delta_c(z)D(0)/D(z)$.

\section[]{The Volume Element for $\lambda_0 \neq 0$}

The number of clusters in a given volume element dV centered at
redshift $z$ and mass range dM
is given by the Press-Schechter abundance times the differential volume
\begin{equation}
dN(M) = n(M,z)dVdM = n(M,z)\frac{r^2drd\Omega}{\sqrt{1-kr^2}}dM
\end{equation}
To integrate this function over a range in redshift rather than in physical
distance, we need to find $r(z)$ for a particular cosmology and make the
transformation to the form
\begin{equation}
dN(M) = n(M,z)f(z)dz\,d\Omega\,dM
\end{equation}
For a universe with $\Omega_0 + \lambda_0 = 1$, $k=0$ and $f(z)=r^2dr/dz$.
The function $r(z)$ can be tabulated by integrating along the null geodesic:
\begin{equation}
r = \int_{0}^{r}dr' = \int_{a}^{a_0}\frac{da}{\dot{a}a}
\end{equation}
We can then express the Friedmann equation in terms of today's values of the
cosmological parameters
\begin{equation}
\dot{a} = aH_0\sqrt{\frac{\Omega_0}{a^3}+\lambda_0}
\end{equation}
and insert this into the geodesic integral to get $r(z)$ in a straightforward
manner.  The integral obtained has an exact solution in terms of elliptic
integrals of the first kind, but expressing it in this manner requires the
integral to be performed from $a=0$.  In addition, for $\lambda_0$ greater
than about 0.7, the elliptic integral must be evaluated outside of its defined
domain.  It is computationally more reliable to evaluate the integral
numerically.  For those who are interested, the ``exact'' solution is
\begin{eqnarray*}
\frac{r(z)}{r_0} & = & \ \ F\left(\cos^{-1}\left[\frac{1+(1-\sqrt{3})u_0}
{1+(1+\sqrt{3})u_0}\right],\kappa\right) \\ 
 & & \!\!\mbox{}-F\left(\cos^{-1}\left[\frac
{1+(1-\sqrt{3})u_z}{1+(1+\sqrt{3})u_z}\right],\kappa\right)
\end{eqnarray*}
where $u_0 = (\lambda_0/\Omega_0)^{1/3}$, $u_z = u_0/(1+z)$, $\kappa = 
\sqrt{2+\sqrt{3}}/2$, and $r_0 = c/(H_0\Omega_0^{1/3}\lambda_0^{1/6}
\sqrt[4]{3})$. 
 
\end{document}